\documentclass{article}
\usepackage{emulateapj}
\usepackage{apjfonts}
\usepackage{graphicx}

\newcommand{\ee}[1]{\times 10^{#1}}
\newcommand{\myunit}[1]{\mbox{$\,\mathrm{#1}$}}

\newcommand{\gram}{\myunit{g}}	
\newcommand{\cm}{\myunit{cm}}	
\newcommand{\second}{\myunit{s}} 
\newcommand{\K}{\myunit{K}}	
\newcommand{\GramPerCc}{\gram\cm^{-3}} 
\newcommand{\erg}{\myunit{erg}} 
\newcommand{\gauss}{\myunit{G}} 

\newcommand{\MeV}{\myunit{MeV}} 

\newcommand{\Msun}{\mbox{$M_\odot$}}
\newcommand{\kpc}{\myunit{kpc}} 

\newcommand{\yr}{\myunit{yr}}	
\newcommand{\km}{\myunit{km}}	
\newcommand{\Hz}{\myunit{Hz}}	

\newcommand{\nuspin}{\nu_{\rm spin}}
\newcommand{\Mspyr}{\mbox{$\,\Msun\yr^{-1}$}} 
\newcommand{\kHz}{\mbox{$\,\mathrm{kHz}$}}
\newcommand{\timdot}{\langle\dot{M}\rangle} 
\newcommand{\tiwd}{\langle W_d \rangle}	
\newcommand{\enu}{\epsilon_\nu}	

\newcommand{\ePhi}{e^{\Phi/c^2}}	
\newcommand{\ePhisq}{e^{2\Phi/c^2}} 

\newcommand{\avmod}{AV18+$\delta$v+UIX*}

\newcommand{\observatory}[1]{\emph{#1}}
\newcommand{\asca}{\observatory{ASCA}}

\newcommand{\axaf}{\chandra}
\newcommand{\beppo}{\observatory{BeppoSAX}}
\newcommand{\chandra}{\observatory{Chandra}}

\newcommand{\rxte}{\observatory{RXTE}}
\newcommand{\xmm}{\observatory{XMM}}

\begin{document}

%
%

\submitted{Submitted to Apj September 16, 1999}

\title{
   Constraints on the Steady-State R-mode Amplitude in Neutron Star
   Transients
}
\lefthead{Steady-state R-modes in Neutron Star Transients}

\author{
   Edward F. Brown\altaffilmark{1} and Greg Ushomirsky
}
\righthead{Brown and Ushomirsky}

\affil{
   Department of Physics and Department of Astronomy
}
\affil{
   601 Campbell Hall, University of California,
   Berkeley, CA 94720--3411;\\
   ebrown@astron.berekely.edu; gregus@fire.berkeley.edu
}

\altaffiltext{1}{
   Current Address: University of Chicago, Enrico Fermi Institute, 5640
   S Ellis Ave, Chicago, IL 60637; brown@oddjob.uchicago.edu 
}



\begin{abstract}
   Recent observations suggest that neutron stars in low-mass X-ray
   binaries rotate within a narrow range of spin frequencies clustered
   around $300\Hz$.  A proposed explanation for this remarkable fact is
   that gravitational radiation from a steady-state r-mode oscillation
   in the neutron star's core halts the spin-up due to accretion.  For
   the neutron star transients, balancing the time-averaged accretion
   torque with gravitational wave emission from steady-state, constant
   amplitude r-mode pulsations implies a quiescent luminosity too bright
   to be consistent with observations (in particular of Aql~X-1).  The
   viscous dissipation (roughly $10\MeV$ per accreted nucleon for a spin
   of $300\Hz$) from such an r-mode makes the core sufficiently hot to
   power a thermal luminosity $\sim 10^{34}\erg\second^{-1}$ when
   accretion halts.  This is the \emph{minimum} quiescent luminosity
   that the neutron star must emit when viscous heating in the core is
   balanced by radiative cooling from the surface, as is the case when
   the core of the star is superfluid.  We therefore conclude that
   either the accretion torque is much less than $\dot{M}(GMR)^{1/2}$,
   or that a steady-state r-mode does not limit the spin rate of the
   neutron star transients.  Future observations with \axaf\ and \xmm\
   promise to further constrain the amount of viscous dissipation in the
   neutron star core.
\end{abstract}

\keywords{
   accretion, accretion disks --- stars: individual(Aquila~X-1) ---
   stars: neutron --- stars: oscillations --- stars: rotation ---
   X-rays: stars --- gravitation --- gravitational waves
}


\section{Introduction}
\label{sec:introduction}

With the launch of \rxte, precision timing of accreting neutron stars
has opened new threads of inquiry into the behavior and lives of these
objects.  The neutron stars in low-mass X-ray binaries (LMXBs) have long
been thought to be the progenitors of millisecond pulsars (see
\markcite{bhattacharya95}{Bhattacharya} 1995 for a review), and a long-standing
observational goal has been the detection of a spin period of a neutron
star in an LMXB. Recent observations (see \markcite{klis99:_millis}{van der Klis} 1999 for a
review) have finally provided conclusive evidence of millisecond spin
periods of neutron stars in about one-third of known Galactic LMXBs.
Altogether, there are seven neutron stars in LMXBs with spin periods
firmly established by either pulsations in the persistent emission (in
the millisecond X-ray pulsar SAX~J1808.4-3658; \markcite{wijnands98}{Wijnands} \& {van der Klis} 1998) or
oscillations during type I X-ray bursts (so-called burst QPOs, first
discovered in 4U~1728--34;
\markcite{strohmayer96:_millis_x_ray_variab_accret}{Strohmayer} {et~al.} 1996).  There are an
additional thirteen sources with twin kHz QPOs for which the neutron
star's spin may be approximately equal to the frequency difference
\markcite{klis99:_millis}({van der Klis} 1999).  A striking feature of all these neutron stars
is that their spin frequencies lie within a narrow range,
$260\Hz<\nuspin<589\Hz$.  The frequency range might be even narrower if
the burst QPOs seen in KS~1731--260, MXB~1743--29, and Aql~X-1 are at
the first harmonic of the spin frequency, as is the case with the
$581\Hz$ burst oscillations in 4U~1636--536
\markcite{miller99:_eviden_antip_hot_spots_durin}({Miller} 1999).  If this is the case,
then the range of observed frequencies is $260\Hz<\nuspin<401\Hz$.  The
neutron stars in LMXBs accrete at diverse rates, from $10^{-11}\Mspyr$
to the Eddington limit, $10^{-8}\Mspyr$. Since disk accretion exerts a
substantial torque on the neutron star and these systems are very old
\markcite{vanParadijs95:LMXB_distrib}({van Paradijs} \& {White} 1995), it is remarkable that these neutron
stars' spins are so tightly correlated, and that none of the neutron
stars are rotating anywhere near the breakup frequency of roughly
$1\kHz$.

Observations therefore suggest that neutron stars in LMXBs are somehow
stuck within a narrow band of spin frequencies well below breakup.  Two
explanations for this convergence of spin frequencies have been
proffered.  \markcite{white97}{White} \& {Zhang} (1997) argued that the magnetospheric spin
equilibrium model (see \markcite{ghosh79:_accret_torque}{Ghosh} \& {Lamb} 1979 and references
therein), which is applicable to the accreting X-ray pulsars, is also at
work in LMXBs.  In this scenario, the neutron star's magnetic field
($B\sim10^9\gauss$) dominates accretion near the stellar surface, and
the Keplerian period at the magnetospheric radius roughly equals the
spin period, so that the accretion stream exerts no net torque on the
star.  Because the sources' luminosities (and presumably accretion
rates) vary by several orders of magnitude, \markcite{white97}{White} \& {Zhang} (1997) noted that
this explanation requires either that the accretion rate be tightly
correlated with the neutron star's magnetic field,
$B\propto\dot{M}^{1/2}$, or that the torque be roughly independent of
accretion rate when the magnetospheric radius approaches the radius of
the neutron star.  Moreover, the persistent pulses typical of magnetic
accretors must also be hidden most of the time.

The other class of theories, first considered by
\markcite{Papaloizou78:gravity_waves}{Papaloizou} \& {Pringle} (1978) and \markcite{Wagoner84}{Wagoner} (1984), invoke the
emission of gravitational radiation to balance the torque supplied by
accretion.  \markcite{bildsten98:gravity-wave}{Bildsten} (1998) proposed that equilibrium
between the accretion torque and gravitational radiation can explain the
narrow range of observed spin frequencies.  The source for the
gravitational radiation could be a mass quadrupole formed by misaligned
electron capture layers in the neutron star's crust
\markcite{bildsten98:gravity-wave}({Bildsten} 1998).  Alternatively, as proposed
independently by \markcite{bildsten98:gravity-wave}{Bildsten} (1998) and
\markcite{andersson99}Andersson, Kokkotas, \&  Stergioulas (1999), current quadrupole radiation from an unstable
r-mode oscillation \markcite{andersson98:_new_class,
friedman98:_axial_instab}({Andersson} 1998; {Friedman} \& {Morsink} 1998) in the liquid core of the neutron star could
also limit the spin, as might occur in hot, newly born neutron stars
(\markcite{lindblom98:_gravit_radiat_instab}Lindblom, Owen, \&  Morsink 1998; \markcite{owen98:_gravit}Owen {et~al.} 1998;
\markcite{andersson99:_gravit_radiat}{Andersson}, {Kokkotas}, \&  {Schutz} 1999).  Because the accretion rate of
LMXBs does vary by several orders of magnitude, the small range of
$\nuspin$ among these objects also requires a correlation between the
quadrupole moment and accretion rate.  This correlation is much less
restrictive, however, than for magnetic equilibrium theories because of
the steep dependence of gravitational wave torque on the spin frequency.

These theories have renewed interest in accreting neutron stars as
gravitational wave sources.  If gravitational radiation does in fact
halt the spin-up of accreting neutron stars, then, regardless of the
mechanism producing the gravitational radiation, the brightest LMXBs
(such as Sco~X-1, with dimensionless strain $h_c\gtrsim2\times10^{-26}$;
\markcite{bildsten98:gravity-wave}{Bildsten} 1998) are also promising sources for
ground-based gravitational wave interferometers, such as LIGO, VIRGO,
GEO, and TAMA \markcite{bildsten98:gravity-wave,Brady99}({Bildsten} 1998; {Brady} \& {Creighton} 1999).  It is not
certain, however, that accreting neutron stars in LMXBs do emit
gravitational radiation.  The \emph{only} evidence to date is their
narrow range of spin frequencies.  It is therefore important to look for
astronomical observations, doable today, that can either corroborate or
rule out the various mechanisms for gravitational radiation from LMXBs.

In this paper we present a new observational test for r-mode driven
gravitational radiation from neutron stars in one set of LMXBs, the
soft X-ray transients.  These are LMXBs in which accretion outbursts,
lasting for days to months, are followed by periods of quiescence,
lasting on the order of years to decades.  Typical time-averaged (over
the recurrence interval, rather than just over the outburst) accretion
rates $\timdot$ for these sources are $\lesssim10^{-10} \Mspyr$,
smaller than those in the brighter persistently accreting LMXBs.  We
show that the quiescent X-ray luminosities of these neutron star
transients (in particular Aql~X-1, which exhibits burst QPOs with a
frequency $549\Hz$; \markcite{zhang98a}{Zhang} {et~al.} 1998) can be used to determine
whether r-modes with amplitudes sufficient to balance the accretion
torque are present in their cores.

Recent theoretical \markcite{brown98:transients}({Brown}, {Bildsten}, \&  {Rutledge} 1998) and observational
\markcite{rutledge99:_therm_x_ray_spect_centaur}({Rutledge} {et~al.} 1999b) works suggest that at
least some fraction of the quiescent luminosity of a neutron star
transient is thermal emission from the neutron star's surface.
Motivated by the possibility of indirectly measuring the core
temperature of an accreting neutron star, we consider the amount of heat
that must be lost, on average, by the neutron star to maintain a thermal
steady state.  If the spins of neutron star transients are set by the
equilibrium between the \emph{time-averaged} accretion torque and
gravitational wave emission by \emph{steady-state} (i.e., constant
amplitude) r-mode pulsations in their cores, the required amplitude of
the pulsations can be computed (\S~\ref{sec:Heat-Visc-Diss}). The
steady-state assumption implies a certain magnitude of viscous
dissipation, i.e., heat deposited directly into the core of the neutron
star.  If the core is superfluid, Urca neutrino emission is suppressed
and this heat escapes as thermal radiation from the surface of the star.
We show (\S~\ref{sec:Observ-sign}) that in this case the X-ray
luminosity in quiescence, $L_q$, would be about 5--10 times greater than
that observed.  If the nucleons in the core are normal, then, as shown
by \markcite{levin99}{Levin} (1999), r-mode pulsations are thermally unstable (at least
for saturation amplitudes of order unity).  In this case it is unlikely
that r-modes are currently excited in any of the known Galactic LMXBs.
If for some reason a thermal steady state could be achieved in a normal
fluid core, however, then Urca neutrino emission would carry away most
of the r-mode heating, and the resulting lower quiescent thermal
luminosities would be consistent, within uncertainties, with
observations.  Our test does not depend on how the r-mode is damped, but
only on the assumptions that the dissipated energy is deposited into the
thermal bath of the star and that the star has reached a rotational and
thermal steady state.  We are only inquiring into total energetics,
i.e., whether the viscous heating present matches that required by the
spin equilibrium with the accretion torque.

\section{R-mode Viscous Heating of Accreting Neutron Stars}
\label{sec:Heat-Visc-Diss}

Recently \markcite{andersson98:_new_class}{Andersson} (1998) and
\markcite{friedman98:_axial_instab}{Friedman} \& {Morsink} (1998) showed that gravitational radiation
excites the r-modes (large scale toroidal fluid oscillations similar to
geophysical Rossby waves) of rotating, \emph{inviscid} stars.
\markcite{lindblom98:_gravit_radiat_instab}Lindblom {et~al.} (1998) compared the gravitational wave
growth timescale $\tau_\mathrm{gr}$ for the r-modes with the viscous
damping timescale $\tau_{v}$ set by shear and bulk viscosities for
normal fluids (i.e., no superfluidity); at rotation rates
$\Omega\lesssim 0.065\Omega_K$, where $\Omega_K=(GM/R^3)^{1/2}$ is the
Keplerian angular velocity at the surface of the star, the damping is
sufficient to preclude unstable growth.  The modes are excited, however,
over a wide range of spin frequencies and temperatures that includes
typical values for the neutron star transients.  

Gravitational waves radiate away angular momentum at a rate
\begin{equation}
   \left.\frac{dJ}{dt}\right|_\mathrm{gr}=-\frac{2J_c}{\tau_\mathrm{gr}},
\end{equation}
where 
\begin{equation}
   J_c=-\frac{3}{2}\alpha^2 \Omega \tilde{J} MR^2
\end{equation}
is the canonical angular momentum of the $(l=2,m=2)$ r-mode
\markcite{Friedma78:CFS_inst,owen98:_gravit}({Friedman} \& {Schutz} 1978; Owen {et~al.} 1998), $\alpha$ is the dimensionless
amplitude of the mode, and $\tilde{J}$ is a dimensionless constant that
accounts for the distribution of mass in the star
\markcite{owen98:_gravit}(Owen {et~al.} 1998).  The gravitational wave growth time
$\tau_\mathrm{gr}$ is negative, which implies instability.  In a
rotational steady state, this angular momentum loss is balanced by the
accretion torque, $N_\mathrm{accr}$.  For a fiducial torque, we assume
that each accreted particle transfers its Keplerian angular momentum to
the neutron star, with a net accretion torque
$N_\mathrm{accr}=\dot{M}(GMR)^{1/2}$.  Using $\tau_\mathrm{gr}$ as
evaluated by \markcite{lindblom98:_gravit_radiat_instab}Lindblom {et~al.} (1998), we find the
steady-state r-mode amplitude \markcite{bildsten98:gravity-wave,levin99}({Bildsten} 1998; {Levin} 1999),
\begin{equation}
   \alpha_\mathrm{steady}=7.9\times10^{-7}
	\left(\frac{\dot{M}}{10^{-11}\Mspyr}\right)^{1/2}
	\left(\frac{300\Hz}{\nuspin}\right)^{7/2},
\end{equation}
such that the fiducial accretion torque $N_\mathrm{accr}$ is balanced by
r-mode angular momentum loss $dJ/dt|_\mathrm{gr}$.

The gravitational radiation reaction adds energy to the unstable r-mode
at a rate
\begin{equation}
   \left.\frac{dE_c}{dt}\right|_\mathrm{gr}=-\frac{2E_c}{\tau_\mathrm{gr}},
\end{equation}
where 
\begin{equation}
   E_c=\frac{1}{2} \alpha^2 \Omega^2 \tilde{J} MR^2
\end{equation}
is the canonical energy of the $(l=2,m=2)$ r-mode
\markcite{Friedma78:CFS_inst,owen98:_gravit}({Friedman} \& {Schutz} 1978; Owen {et~al.} 1998).  In a steady state all of this
energy must be dissipated by viscous processes at a rate $W_d=dE_c/dt$.
In terms of the accretion luminosity, 
$L_A=GM\dot{M}/R=N_\mathrm{accr}\Omega_K$, the dissipation rate is
\begin{equation}\label{eq:newtonian-dissipation}
   \frac{W_d}{L_A}=-\frac{1}{\Omega_K}
	\frac{dE_c/dt|_\mathrm{gr}}{dJ_c/dt|_\mathrm{gr}}=
	\frac{1}{3}\frac{\Omega}{\Omega_K}.
\end{equation}

The viscosity in the neutron star originates from several possible
sources.  For normal $npe$ matter, calculations of the viscous transport
coefficients exist only at near-nuclear densities \markcite{flowers79}(Flowers \& Itoh 1979).
The components of such a core are strongly degenerate, and phase-space
restrictions impart a characteristic $T^{-2}$ dependence to the shear
viscosity \markcite{cutler87}({Cutler} \& {Lindblom} 1987).  Compressing a fluid element of neutron star
matter causes it to emit neutrinos as the $npe$ mixture reestablishes
$\beta$-equilibrium, so the bulk viscosity has an Urca-like $T^6$
dependence \markcite{sawyer89:_bulk}(Sawyer 1989).  Another possibility for the
viscosity is that it is caused by mutual friction in the neutron-proton
superfluid \markcite{mendell91:_super_hydrod_rotat_neutr_stars}({Mendell} 1991).  In this
case the viscous damping is independent of temperature.

While the total amount of viscous dissipation $W_d$ depends only on the
assumption of a steady-state r-mode amplitude, the amount of heat
actually deposited into the star depends on the nature of the damping.
If the dominant viscous mechanism is bulk viscosity (i.e., for core
temperatures $T\gtrsim10^9 \K$), then the dissipated energy is released
in the form of neutrinos, which promptly leave the star.  The core
temperatures of LMXBs are most likely less than $10^9 \K$, however, in
which case the dissipation mechanism is either shear viscosity or mutual
friction.  For both of these mechanisms, the heat $W_d$ is deposited
directly into the core of the star; we shall assume this to be the case
in the rest of this paper.

\markcite{levin99}{Levin} (1999) first noted that, if the nucleons in the core are
normal, the r-modes damped by shear viscosity are likely to be thermally
unstable, at least for saturation amplitudes of order unity.  The
heating from the shearing motions decreases the viscosity, and so the
r-mode amplitude increases, which heats the star even more.  The result
is a thermal and dynamical runaway.  As envisaged by \markcite{levin99}{Levin} (1999),
the neutron star enters a limit cycle of slow spin-up to some critical
frequency, at which the r-mode becomes unstable, followed by a rapid
spin-down until the mode is once again damped.  As the neutron star
cools, accretion again exerts a positive torque on the star, and the
cycle repeats.  Because the r-modes are present at a nonzero amplitude
for only $\sim10^{-7}$ of the entire cycle's duration, it is unlikely
that any of the known LMXBs harbor active r-modes \emph{and} have normal
fluid cores.

For a superfluid core, where the damping is due to mutual friction (and
hence independent of temperature), the neutron star can reach a state of
three-fold equilibrium \markcite{bildsten98:gravity-wave,levin99}({Bildsten} 1998; {Levin} 1999): the
temperature is set by the balance of viscous heating and radiative or
neutrino cooling, the r-mode's amplitude is set by the balance of
gravitational radiation back-reaction and viscous damping, and the spin
is set by the balance of accretion torque and angular momentum loss to
gravitational radiation.  It is this scenario that we shall examine for
existing evidence of r-mode spin regulation.

While a neutron star accretes, its luminosity is dominated by the
release of the infalling matter's gravitational potential energy,
$L_A\approx190\MeV(\dot{M}/m_b)$, where $m_b$ is the average nucleon
mass.  Nuclear burning (either steady or via type~I X-ray bursts) of the
accreted hydrogen and helium generates an additional $\sim5\MeV$ per
accreted nucleon.  Most of this heat is promptly radiated away, however,
and no more than a few percent diffuses inward to heat the interior
\markcite{fujimoto84,fujimoto87}({Fujimoto} {et~al.} 1984, 1987).  Nuclear reactions in the deep crust (at
$\rho\gtrsim5\ee{11}\GramPerCc$) release about $1\MeV$ per accreted
nucleon \markcite{sato79,blaes90:_slowl,haensel90a}({Sato} 1979; {Blaes} {et~al.} 1990; {Haensel} \& {Zdunik} 1990) and heat the crust
directly \markcite{brown98a,Brown99}({Brown} \& {Bildsten} 1998; {Brown} 2000).

In addition to the crustal reactions, the viscous dissipation of r-modes
constitutes another heat source in the neutron star's core.  For a
fiducial neutron star with $M=1.4\Msun$ and $R=10\km$,
equation~(\ref{eq:newtonian-dissipation}) implies that $W_d/L_A =
0.046(\nuspin/300\Hz)$, or
\begin{equation}\label{eq:Wd-mag}
   W_d\approx 8.9 \MeV \left(\frac{\dot{M}}{m_b}\right)
      \left(\frac{\nuspin}{300\Hz}\right).
\end{equation}
This heating is very substantial, as it is much greater than the amount
of nuclear heating from the crustal reactions.  The prospects for
detecting the effect of core r-mode heating in \emph{steadily} accreting
neutron stars are dim, unfortunately, as it is dwarfed by the accretion
luminosity (which is a factor of 20 brighter).  The thermal emission
from the neutron star is directly observable, however, if accretion
periodically halts, as in the neutron star transients (the cooling
timescale of the heated core is $\sim 10^4\yr$).  While continued
accretion at low levels between outbursts may contribute some of the
quiescent luminosity (see \markcite{brown98:transients}{Brown} {et~al.} 1998 for a
discussion), the thermal emission from the hot crust of the neutron star
is impossible to hide, and so observations of $L_q$ set an upper limit
on the core temperature.  The neutron stars in soft X-ray transients
therefore offer the best prospects to look for evidence of viscous
heating.  In the next section we predict the quiescent luminosity $L_q$
that arises because of the r-mode heating and compare it to the observed
luminosities of several neutron star transients.


\section{The Quiescent Luminosities of Neutron Star Transients}
\label{sec:Observ-sign}

The neutron star accretes fitfully, so the spin period and the r-mode
amplitude oscillate about the equilibrium defined by the time-averaged
accretion rate, $\timdot\equiv t_r^{-1}\int \dot{M}\,dt$, where $t_r$
is the recurrence interval.  Moreover, the timescale for viscous
dissipation to heat the core is
\begin{equation}
   t_H \sim \frac{c_p T}{\tiwd} \frac{M}{m_b}
	\approx 6\times10^4\yr
	\left(\frac{10^{-11}\Mspyr}{\timdot}\right),
\end{equation}
where $c_p$ is the specific heat per baryon and $\tiwd$ is the viscous
heating averaged over an outburst/quiescent cycle.  Because $t_H$ is
much longer than the outburst recurrence time (typically of order years
to decades), the core should remain fixed at the temperature set by the
balance (over many outburst/quiescent cycles) between heating and
cooling processes.  We may therefore compute the viscous dissipation
using $\timdot$.  Some simple estimates of the equilibrium core
temperatures and the resulting quiescent luminosities, for when both
radiative and neutrino cooling are important, are presented first
(\S~\ref{sec:simple-estimates}).  This is followed, in
\S~\ref{sec:Numer-calc}, by detailed numerical calculations of the
neutron star's thermal structure and a comparison
(\S~\ref{sec:comp-observ-trans}) to observations of several neutron star
transients.

\subsection{Simple Estimates}
\label{sec:simple-estimates}

In a thermal steady state, the neutron star interior is cooled both by
neutrinos emitted from the core and crust and by photons emitted from
the surface.  To begin, we estimate the luminosity and the equilibrium
core temperature set by balancing the heat deposited during an
outburst/recurrence cycle, $\tiwd$, with each cooling mechanism
individually.  First, if neutrino emission from the core is negligible
(e.g., if the core is superfluid and the Urca processes are
exponentially suppressed), then all of the heat generated by viscous
dissipation, $\tiwd$, is conducted to the surface of the neutron star
and escapes as thermal radiation during quiescence.  For the interior to
be in a thermal steady state, the quiescent luminosity must then be
\begin{equation}
\label{eq:rad_balance}
   L_q\approx \tiwd = 5.4\ee{33}\erg\second^{-1}
   \left(\frac{\timdot}{10^{-11}\Mspyr}\right) 
   \left(\frac{\nuspin}{300 \Hz}\right).
\end{equation}
This estimate depends only on the assumption that neutrino emission is
suppressed, and is independent of the crust microphysics.

As a check, we estimate the temperature of the neutron star core.  In
quiescence the atmosphere and crust come to resemble a cooling neutron
star \markcite{bildsten97a,brown98:transients}({Bildsten} \& {Brown} 1997; {Brown} {et~al.} 1998).  For the temperature
increase through the atmosphere and upper crust, we use the fit of
\markcite{gudmundsson83}{Gudmundsson}, {Pethick}, \&  {Epstein} (1983),
\begin{equation}
\label{eq:GPE_atm}
   L_\gamma\approx 8.2\ee{32}\erg\second^{-1}
      \left(\frac{T_b}{10^8\K}\right)^{2.2},
\end{equation}
where $T_b$ is the temperature at a fiducial boundary
$\rho_b=10^{10}\GramPerCc$.  Equating $L_\gamma$ with $L_q$ from
equation~(\ref{eq:rad_balance}) gives an estimate of the temperature in
the upper crust,
\begin{equation}\label{eq:Tb-radiative}
   T_b \approx 2.4\ee{8}\K
   \left(\frac{\timdot}{10^{-11}\Mspyr}\right)^{0.45}
   \left(\frac{\nuspin}{300\Hz}\right)^{0.45}.
\end{equation}
To relate $T_b$ to the core temperature $T_c$, we use approximate
analytic expressions (\markcite{Brown99}{Brown} 2000; eqs.~[22] and [23]) for the
crust temperature to obtain
\begin{equation}\label{eq:Tc-to-Tb}
\left(\frac{T_c}{10^8\K}\right)^2\approx\left(\frac{T_b}{10^8\K}\right)^2+
	4.9\left(\frac{L_q}{10^{34}\erg\sec^{-1}}\right),
\end{equation}
where we have neglected the luminosity due to crustal nuclear reactions.
Substituting from equation~(\ref{eq:Tb-radiative}) for $T_b$, we obtain
the core temperature in the absence of neutrino emission,
\begin{equation}\label{eq:Tc-radiative}
   T_c \approx 2.9\ee{8}\K
   \left(\frac{\timdot}{10^{-11}\Mspyr}\right)^{0.45}
   \left(\frac{\nuspin}{300\Hz}\right)^{0.45},
\end{equation}
where the scalings for $\timdot$ and $\nuspin$ are obtained by dropping
the second term on the right in equation~(\ref{eq:Tc-to-Tb}). This
estimate agrees quite well with the detailed calculations described in
\S~\ref{sec:Numer-calc}. 

The core neutrino emissivity is, for modified Urca processes
\markcite{shapiro83}(Shapiro \& Teukolsky 1983), $L_\nu^\mathrm{Urca}\approx 7.4\ee{31}
(T_c/10^8\K)^8$, multiplied by a superfluid reduction factor that goes
roughly as $\exp(-\Delta/kT_c)$, where $\Delta$ is the superfluid gap
energy \markcite{yakovlev95}(Yakovlev \& Levenfish 1995).  For $\Delta>kT_c$ the net Urca neutrino
luminosity is much less than $L_q$, so that
equation~(\ref{eq:rad_balance}) is self-consistent.  Neutrino emission
from crust neutrino bremsstrahlung \markcite{kaminker99:_neutr}(Kaminker {et~al.} 1999) at the
temperature $T_b$ (eq.~[\ref{eq:Tb-radiative}]) is also not significant,
although at higher $\timdot$ it is competitive with radiative
cooling. Hence for accretion rates typical of neutron star transients,
the majority of the deposited heat is conducted to the surface, and
equation~(\ref{eq:rad_balance}) provides a robust estimate of the
radiative luminosity of the star.
%
%
%

Alternatively, if core neutrino emission is not suppressed (i.e., the
nucleons are not superfluid), then modified Urca processes are the
dominant coolant and $L_\nu^\mathrm{Urca}\approx W_d$.  In this case the
core temperature is
\begin{equation}
T_c \approx 1.7\ee{8}\K
	\left(\frac{\timdot}{10^{-11}\Mspyr}\right)^{1/8} 
	\left(\frac{\nuspin}{300\Hz}\right)^{1/8},
\end{equation}
and is smaller than if the core were superfluid.  A colder core implies
a dimmer thermal luminosity from the surface.  In order to estimate
$L_q$, we write $W_d=L_q+L_\nu^\mathrm{Urca}(T_c)$, where
$L_q=L_\gamma(T_b)$, and $T_b$ is related to $T_c$ by
equation~(\ref{eq:Tc-to-Tb}).  Under the assumption that $L_q\ll
L_\nu^\mathrm{Urca}$, the solution of the resulting transcendental
equation is
\begin{equation}
   L_q \approx 1.8\ee{33}\erg\second^{-1}
	\left(\frac{\timdot}{10^{-11}\Mspyr}\right)^{0.3}
	\left(\frac{\nuspin}{300 \Hz}\right)^{0.3},
\end{equation}
where we obtain the scalings for $\timdot$ and $\nuspin$ by dropping the
second term on the right-hand side of equation~(\ref{eq:Tc-to-Tb}).  In
this case $L_q$ is less than $W_d$ and $L_\nu^\mathrm{Urca}$, so our
assumption that core neutrino emission is the dominant coolant is
self-consistent.


These estimates neglect cooling from other neutrino-producing
mechanisms, such as neutrino bremsstrahlung in the crust.  Moreover, the
core neutrino emissivity depends on the local proper temperature, which
increases towards the center of the star because of the gravitational
redshift.  We now describe our detailed calculations, which take these
effects into account.


\subsection{Numerical Calculations}
\label{sec:Numer-calc}

To calculate the expected quiescent luminosities of accreting neutron
stars, we compute hydrostatic neutron star models by integrating the
post-Newtonian stellar structure equations \markcite{thorne77}(Thorne 1977) for the
radius $r$, gravitational mass $m$, potential, and pressure with the
equation of state \avmod\ \markcite{akmal98}(Akmal, Pandharipande, \& Ravenhall 1998), as described in
\markcite{Brown99}{Brown} (2000).  With the hydrostatic structure specified, the
luminosity $L$ and temperature $T$ are found by solving the entropy and
flux equations \markcite{thorne77}(Thorne 1977),
\begin{eqnarray}
\label{eq:entropy}
   e^{-2\Phi/c^2} \frac{\partial}{\partial r}\left(L \ePhisq\right)
      - 4\pi r^2 n \left(\epsilon_r - \enu\right)
      \left(1-\frac{2Gm}{rc^2}\right)^{-1/2} &=& 0\\
\label{eq:flux}
   e^{-\Phi/c^2} K \frac{\partial}{\partial r}\left(T \ePhi\right) 
      + \frac{L}{4\pi r^2}\left(1-\frac{2Gm}{rc^2}\right)^{-1/2} &=& 0.
\end{eqnarray}
Here $\epsilon_r$ and $\enu$ are the nuclear heating and neutrino
emissivity per baryon, $n$ is the baryon density, and $K$ is the thermal
conductivity.  The potential $\Phi$ appears in the time-time component
of the metric as $e^{\Phi/c^2}$ (it governs the redshift of photons and
neutrinos; \markcite{misner73:_gravit}Misner, Thorne, \& Wheeler 1973).  We neglect in
equation~(\ref{eq:entropy}) terms arising from compressional heating, as
they are of order $T\Delta s (\dot{M}/M)$ \markcite{fujimoto82:_helium}({Fujimoto} \& {Sugimoto} 1982),
$s$ being the specific entropy, and are negligible throughout the
degenerate crust and core \markcite{brown98a}({Brown} \& {Bildsten} 1998).  We do not include heating
from nuclear reactions in the deep crust. This has the effect of
underestimating slightly (by $\lesssim 10\%$) the quiescent luminosity
of the neutron star transient.  Equations~(\ref{eq:entropy})
and~(\ref{eq:flux}) are integrated outwards to a density $\rho_b =
10^{10}\GramPerCc$.  We there impose a boundary condition relating $L$
and $T$ with the fitting formula of \markcite{potekhin97}{Potekhin}, {Chabrier}, \&  {Yakovlev} (1997) for a partially
accreted crust.  By incorporating a parameter describing the depth of a
light element (H and He) layer, this formula differs from that of
\markcite{gudmundsson83}{Gudmundsson} {et~al.} (1983), which we used for our simple estimates
(\S~\ref{sec:simple-estimates}).  We set the depth of this light element
layer to where the density is $\approx 10^5\GramPerCc$, which is roughly
where the accreted material burns to heavier elements \markcite{hanawa86}({Hanawa} \& {Fujimoto} 1986).

The high thermal conductivity of the neutron star's core insures that it
is very nearly isothermal, regardless of the detailed dependence of the
heating rate $\epsilon_r$ on the radius.  For the core temperatures
typical of LMXBs, bulk viscosity is unimportant, so we assume that the
heating is from ordinary shear viscosity.  The rate per unit volume is
just $2\eta\delta\sigma^{ab}\delta\sigma^{*}_{ab}$, where $\eta$ is the
shear viscosity and $\delta\sigma_{ab}$ is the kinematic shear.  If we
neglect the dependence of shear viscosity on density (since the density
is approximately constant in the neutron star's core), this rate is just
proportional to $r^2$ for an $(l=2,m=2)$ r-mode
\markcite{lindblom98:_gravit_radiat_instab}(Lindblom {et~al.} 1998).  Hence we take
$\epsilon_r\propto r^2$, and normalize it so that the heating rate, when
integrated over the core, satisfies
equation~(\ref{eq:newtonian-dissipation}).

The microphysics used to integrate equations~(\ref{eq:entropy}) and
(\ref{eq:flux}) is fully described in \markcite{Brown99}{Brown} (2000), so here we just
highlight two modeling uncertainties.  First, standard calculations
presume that the neutron star's crust is a pure lattice, and hence the
conductivity is dominated by electron-phonon scattering.  Over the
lifetime of an LMXB, however, the neutron star can easily accrete enough
matter to replace its entire crust (requiring about $0.01\Msun$).  The
accreted crust is formed from the products of hydrogen and helium
burning and is likely to be very impure \markcite{schatz99}({Schatz} {et~al.} 1999).  A lower
conductivity from impurities lowers the surface temperature, and hence
the quiescent luminosity, for a given core temperature.  We model the
low thermal conductivity of a very impure crust by using electron-ion
scattering \markcite{haensel96b}({Haensel}, {Kaminker}, \&  {Yakovlev} 1996) throughout the crust.

The second modeling uncertainty is the superfluid transition
temperatures, for which estimates vary widely \markcite{tsuruta98}(see {Tsuruta} 1998, and
references therein).  When the core temperature is much less
than the superfluid transition temperature, emissivity from Cooper
pairing is unimportant \markcite{yakovlev99:_neutr_cooper}({Yakovlev}, {Kaminker}, \&  {Levenfish} 1999) and the
superfluidity suppresses the neutrino emission by roughly the Boltzmann
factor $\exp(-\Delta/k_BT)$, for a superfluid gap energy $\Delta$.  We
perform our calculations for two models, one with superfluidity
parameterized as in \markcite{Brown99}{Brown} (2000), with a typical gap energy
$\Delta\sim0.5\MeV$, and another model with a normal core, $\Delta=0$.

Figure~1 demonstrates the thermal structure of such a
neutron star with a time-averaged accretion rate
$\timdot=2.4\ee{-11}\Mspyr$ (the rate inferred for Aql~X-1) for a spin
frequency of $275\Hz$ (\emph{solid lines}) and $549\Hz$ (\emph{dotted
lines}).  If the core is superfluid (the upper pair of curves), then the
neutrino luminosity from crust bremsstrahlung (region leftward of the
vertical dot-dashed line) is roughly comparable to the photon
luminosity.  In contrast, if the core were normal (so that the modified
Urca processes were unsuppressed) but the viscosity remained independent
of temperature (so that a thermal steady state could be reached), then
only about 10\% of the heat generated in the core would be conducted to
the surface.  The rest of the heat is balanced by modified Urca neutrino
emission.  The core temperature and fraction of viscous heat conducted
to the surface compare well with the estimates in
\S~\ref{sec:simple-estimates}.

\centerline{\includegraphics[width=\hsize]{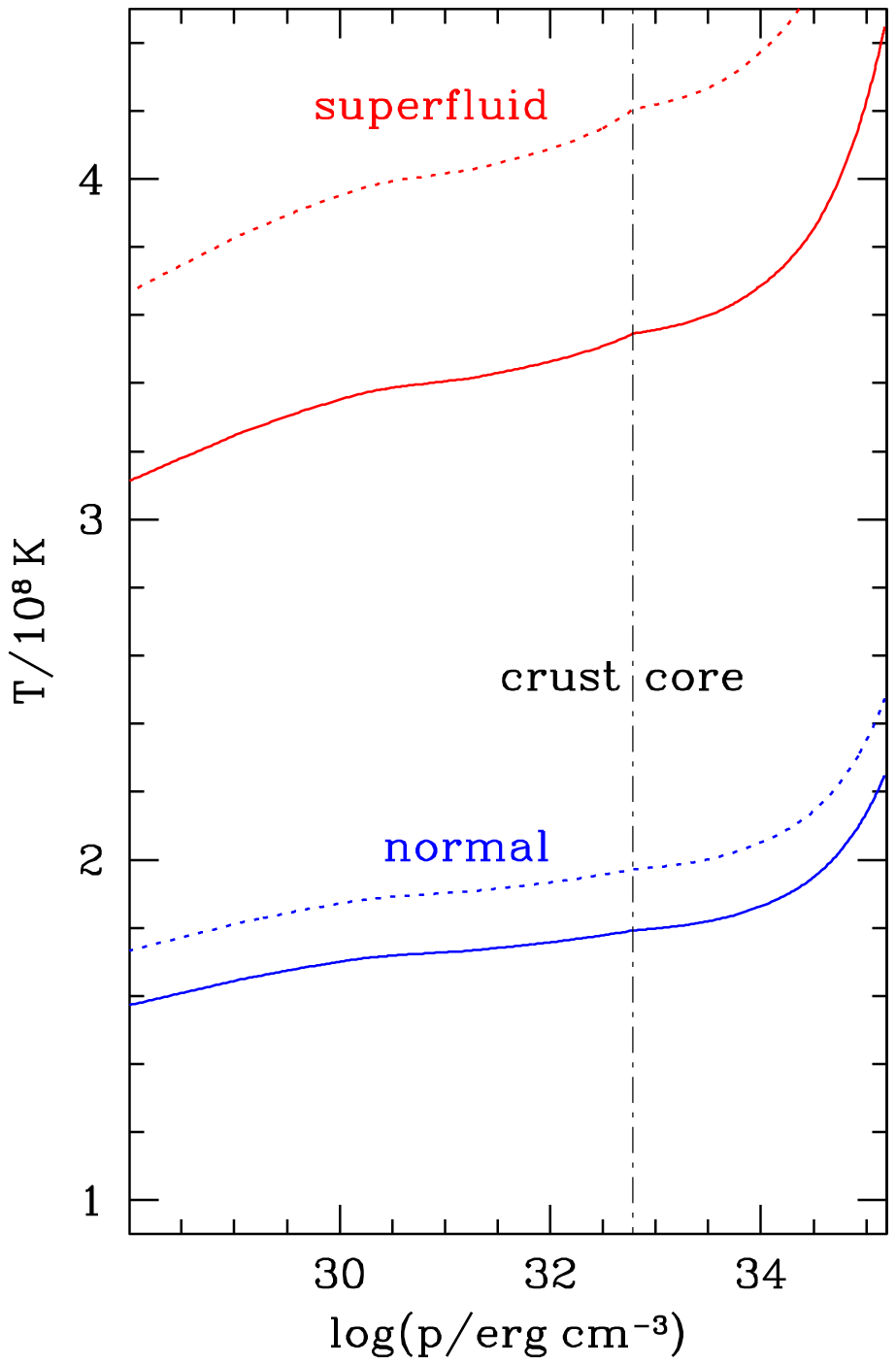}}
\label{fig:aql-sf}
{\footnotesize%
   \textsc{Fig. 1---} The thermal structure of a neutron star accreting
   at a time-averaged rate of $2.4\ee{-11}\Mspyr$ (e.g., Aql~X-1), for
   two different spin frequencies: 275\Hz\ (\emph{solid lines}) and
   549\Hz\ (\emph{dotted lines}).  The upper pair of curves are for a
   superfluid core (region rightward of the vertical dot-dashed line);
   the lower pair, for a normal core.}

\subsection{Comparison to Observed Transients}
\label{sec:comp-observ-trans}

A superfluid core is cooled mainly by conduction of heat to the surface,
at least until the interior temperature is high enough to activate crust
neutrino bremsstrahlung.  Figure~2 shows the expected quiescent
luminosity $L_q$ as a function of $\timdot$ for this case, with a range
(\emph{shaded region}) of rotation frequencies $200\Hz<f<600\Hz$.  The
inferred $\langle\dot{M}\rangle$ and $L_q$ for several neutron star
transients are also plotted (\emph{squares}) for comparison.  With the
exception of EXO~0748-676%
\footnote{%
   EXO~0748-676 is likely to accrete during quiescence, as suggested by
   observations (with \asca) of variability on timescales $\gtrsim
   1000\second$
   \protect\markcite{corbet94:_is_exo,thomas97:_discov_new_soft_x_ray}({Corbet} {et~al.} 1994; {Thomas} {et~al.} 1997).},
the neutron star transients with measured quiescent luminosities are
too dim, by a factor of 5--10, to be consistent with viscous heating
of the magnitude assumed here.  We must conclude, then, that
\emph{either the accretion torque is much less than
$\timdot(GMR)^{1/2}$, or that a steady-state r-mode does not set their
spin.}

The quiescent luminosities for Aql~X-1, Cen~X-4, and 4U~1608--522 use
the bolometric corrections appropriate for a H atmosphere spectrum
\markcite{rutledge99:_prosp_means_distin_between_neutr}({Rutledge} {et~al.} 1999a); $L_q$ for the
Rapid Burster is from \markcite{asai96b}{Asai} {et~al.} (1996a).  We infer the time-averaged
accretion rate from $\timdot\approx (t_o/t_r)(L_o/GMR^{-1})$, where\\
\centerline{\includegraphics[width=\hsize]{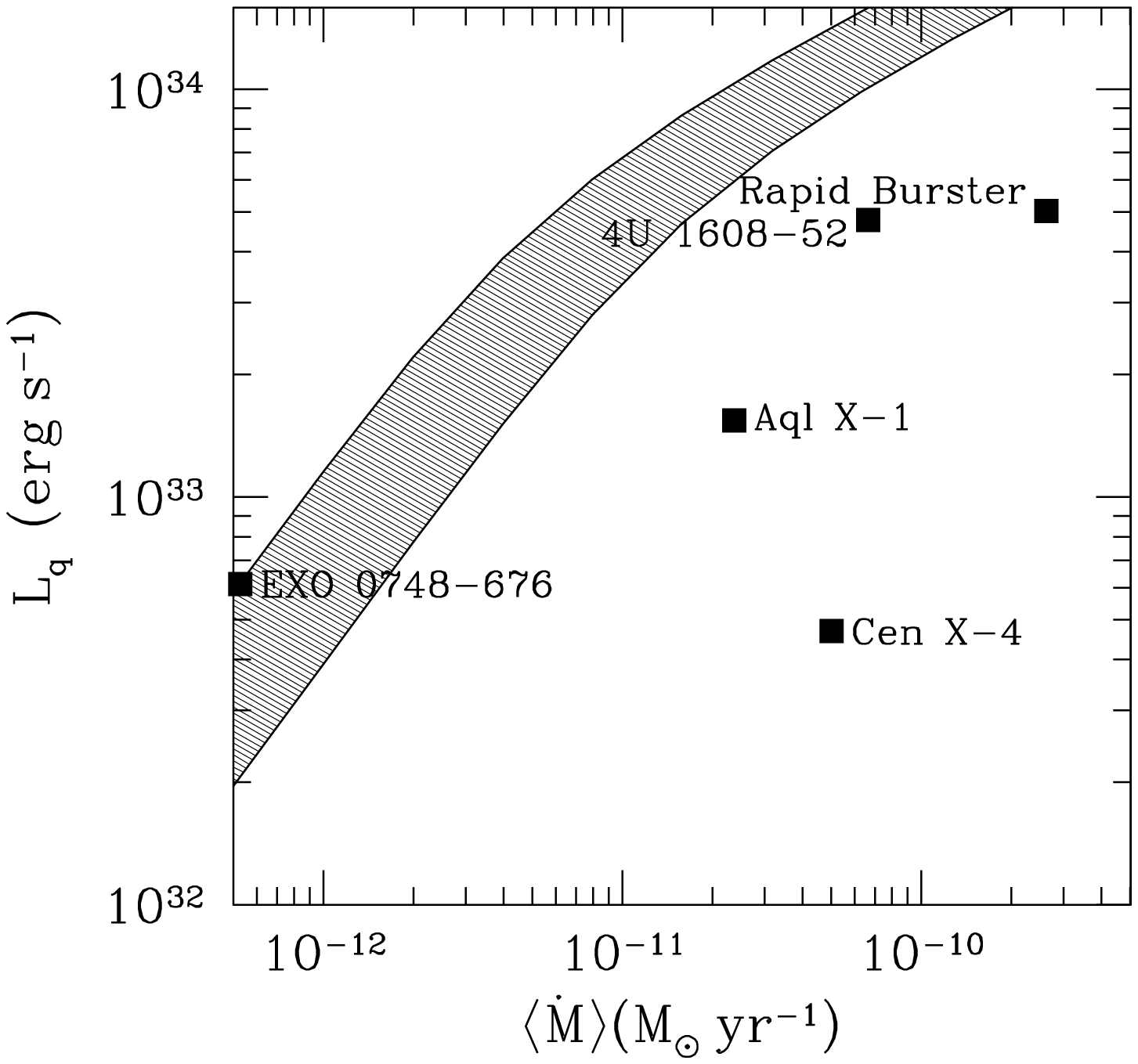}}
\label{fig:rmode-L-SF}{\footnotesize%
   \textsc{Fig 2.---} Quiescent luminosities as a function of
   time-averaged accretion rate $\timdot$.  The heating from viscous
   dissipation of the r-mode is from
   eq.~(\protect\ref{eq:newtonian-dissipation}), and the neutrino
   emission from the core is suppressed by nucleon superfluidity.  The
   shaded region corresponds to rotation frequencies between $200\Hz$
   (\emph{lower curve}) and $600\Hz$ (\emph{upper curve}).  Neutrino
   cooling from crust bremsstrahlung is important rightward (i.e., at
   higher $\timdot$) of the knee in the shaded region.  Also shown are
   the inferred quiescent luminosities and time-averaged accretion rates
   for several neutron star transients.
}
\vspace*{1.0em}\\
$t_o$ and $L_o$ are the outburst duration and luminosity and the distances are
taken from \markcite{chen97}{Chen}, {Shrader}, \& {Livio} (1997).\footnote{%
   Recent observations \markcite{callanan99:_v1333_aquil}({Callanan}, {Filippenko}, \&  {Garcia} 1999) resolved the
   optical counterpart of Aql~X-1 into two objects.  We use the distance
   estimate ($2.5\kpc$) of \markcite{chevalier99:_magnit_aql_x}{Chevalier} {et~al.} (1999), which
   accounts for the interloper star.}
Outburst fluences for Aql~X-1 and the Rapid Burster are accurately
known (\rxte/All-Sky Monitor public data); for the remaining sources
$\timdot$ is estimated from peak luminosities and outburst rise and
decay timescales \markcite{chen97}({Chen} {et~al.} 1997).

Our estimates for $\timdot$ depend on the inferred source distance.
When most of the r-mode heating $W_d$ is conducted to the surface,
however, as in the superfluid core case for $\timdot\lesssim
10^{-11}\Mspyr$, the predicted quiescent luminosity is $L_q\approx
W_d\propto\timdot$ (see eqs.~[\ref{eq:newtonian-dissipation}] and
[\ref{eq:Wd-mag}]), and hence depends on the source distance in the same
way as does $\timdot$.  Therefore, our comparison of $L_q$ predicted
from r-mode heating and the quiescent luminosity actually observed is
\emph{independent} of distance.  In this regime, the relation between
$L_q$ and $\timdot$ \emph{is also independent of the microphysics in the
crust.}

As shown by \markcite{levin99}{Levin} (1999), the temperature dependence of viscosity in
a normal fluid likely prevents a steady-state r-mode.  For comparison,
however, we plot in Figure~3 the case where modified Urca neutrino
emission from the core is allowed (as it would be in a normal fluid) but
the r-mode amplitude is steady, i.e., we assume that a
thermogravitational runaway has somehow been avoided.  As a result,
neutrino emission efficiently cools the core, and so the radiative
luminosity $L_q$ is less for a given $\timdot$.  For this case, with the
exception of Cen~X-4, the neutron star transients have quiescent
luminosities roughly consistent with that predicted.  Because there is a
characteristic core temperature, namely, that at which neutrino cooling
equals radiative cooling, the relation between $\timdot$ and $L_q$ is
no\\
\centerline{\includegraphics[width=\hsize]{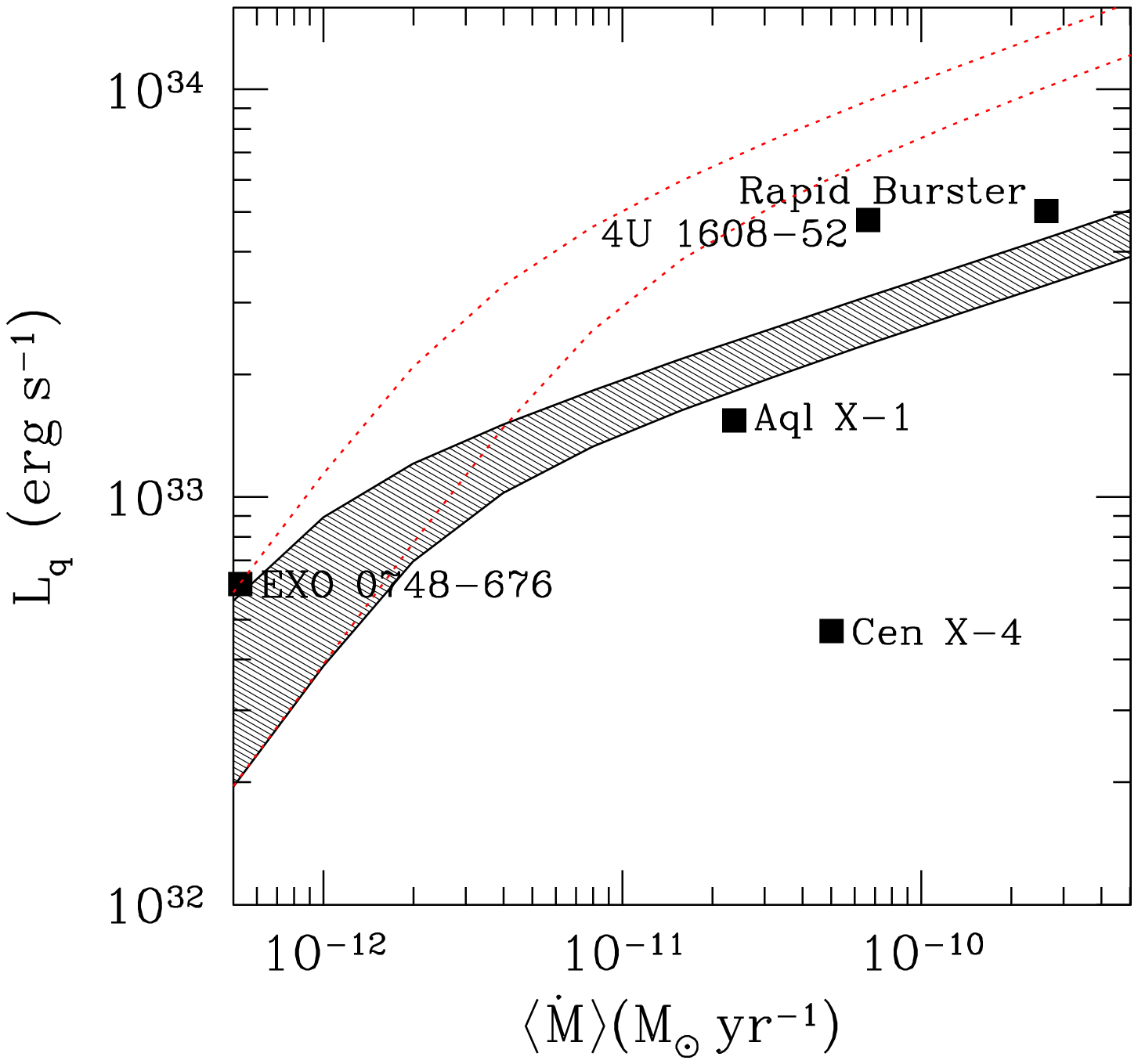}}
\label{fig:rmode-L}{\footnotesize%
   \textsc{Fig. 3---} The same as Fig.~2,
   but for a normal core.  We also show (\emph{thin dotted lines})
   $L_q(\timdot)$ for a neutron star with a crust of light elements at
   $\rho<10^{10}\GramPerCc$. }
\vspace*{1.0em}\\
longer independent of distance, unlike the case shown in Figure~2.  The
knee in the shaded region is where the neutrino and photon luminosities
are comparable.  Rightward of this knee ($\timdot\gtrsim
10^{-12}\Mspyr$) neutrino cooling prevents the core temperature, and
hence the photon luminosity, from rising rapidly with increasing
$\timdot$.  Should the crust have a higher conductivity (e.g., if it
were more pure) than we have assumed here, then the shaded region
rightward of the knee would move upwards, i.e., the predicted $L_q$
would be even higher.  To illustrate this we computed $L_q(\timdot)$
using the $L_\gamma(T_b)$ relation for a crust composed of light
elements (and having a higher conductivity) for densities less than
$\rho_b=10^{10}\GramPerCc$ (\emph{dotted lines}).

It should be noted that the actual thermal radiation from a neutron
star's surface is in general \emph{less} than the observed quiescent
luminosity, since other emission mechanisms are possible, such as
accretion via a low-efficiency advective flow \markcite{narayan96a}({Narayan}, {McClintock}, \&  {Yi} 1996) or
magnetospheric emission \markcite{campana98b}({Campana} {et~al.} 1998a).  Evidence for other,
non-thermal emission processes are the hard power-law tails observed
from Cen~X-4 (\asca; \markcite{asai96a}{Asai} {et~al.} 1996b) and Aql~X-1 (\beppo;
\markcite{campana98a}{Campana} {et~al.} 1998b).  In addition, variability on timescales of a few
days has been observed from Cen~X-4 \markcite{vanParadijs87,campana97}({van Paradijs} {et~al.} 1987; {Campana} {et~al.} 1997).  As
a result, a plot showing thermal emission (as opposed to observed $L_q$)
would have the data points shifted downward in Figures~2 and~3.  In
other words, the quiescent luminosity inferred from observations is
likely to overestimate the actual thermal emission from the neutron
star.  This strengthens our conclusion regarding the incompatibility of
steady-state r-mode heating with the observations.

There are stronger neutrino emission mechanisms possible than modified
Urca and crust bremsstrahlung.  Recently, there has been renewed
interest in the direct Urca process \markcite{lattimer91}({Lattimer} {et~al.} 1991), which is allowed
if the proton fraction exceeds 0.148 or if hyperons are present
\markcite{prakash92:_rapid_delta}({Prakash} {et~al.} 1992).  Other exotic mechanisms may be
possible, including pion condensates \markcite{umeda94}({Umeda} {et~al.} 1994), kaon condensates
\markcite{brown88:_stran}(Brown {et~al.} 1988), or quark matter \markcite{iwamoto82:_neutr}(Iwamoto 1982).  The
exotic mechanisms have the same temperature dependence as the direct
Urca ($\propto T^6$) but are weaker.  Should any of these enhanced
processes occur, the core will be much colder, and the heat radiated
from the surface much weaker, than in the calculations here.  For
example, balancing the viscous heating with neutrino emission from a
pion condensate, $L_\nu^\pi \approx 2.0\ee{39}
(T/10^8\K)^6\erg\second^{-1}$ \markcite{shapiro83}(Shapiro \& Teukolsky 1983), implies that $T_c
\approx 1.2 \times 10^{7} (\timdot/10^{-11}\Mspyr)^{1/6}\K$, and, from
equations~(\ref{eq:GPE_atm}) and (\ref{eq:Tc-to-Tb}), that $L_q\approx
6.0\ee{30}(\timdot/10^{-11}\Mspyr)^{0.45}\erg\second^{-1}$.  This is much
dimmer than that observed.  Of course, it is possible that superfluidity
reduces $L_\nu$ such that the core temperature is just enough to explain
the observed quiescent emission.  It is difficult, however, to arrange
\emph{all} of the sources to obey such a relation.

\section{Conclusions}
\label{sec:conclusions}

Using the assumption that the accretion torque is balanced by angular
momentum loss from gravitational radiation by an r-mode pulsation of
constant amplitude, we find that the expected quiescent luminosities of
the neutron star X-ray transients, for rotation rates of 200--600\Hz,
are characteristically brighter than those observed.  Reconciling the
observations with the presence of r-mode heating requires that neutrino
emission from the core be unsuppressed, as for a normal core.  In this
case, however, the r-mode is thermally unstable and cannot remain at a
constant amplitude, unless some mechanism prevents a runaway.  It
therefore seems unlikely that the spin frequency of Aql~X-1 is a
signature of a steady-state core r-mode pulsation.  We note, however,
that the same conclusion cannot be drawn for the bright, persistent
LMXBs (such as Sco~X-1, which could be detected by gravitational wave
experiments soon to be operational).  Uncertainties in the nuclear
burning and the accretion luminosity cannot constrain the surface
thermal luminosity to within $\lesssim 5\%$, which is necessary to
differentiate the r-mode heating from the accretion luminosity.

In addition to Aql~X-1, there is one other neutron star transient which
is known to be spinning rapidly, and that is the $401\Hz$ accreting
pulsar \markcite{wijnands98}({Wijnands} \& {van der Klis} 1998) in the transient SAX~J1808.4--3658
\markcite{intZand98}({in't Zand} {et~al.} 1998).  This source has not yet been detected in quiescence.
Given a recurrence interval of $1.5\yr$, an outburst duration of
$\approx20$~day, and an outburst accretion rate of
$\approx3\ee{-10}\Mspyr$ \markcite{intZand98}({in't Zand} {et~al.} 1998), we expect a quiescent
luminosity $\gtrsim 5.8\ee{33}\erg\second^{-1}$ if the core is
superfluid and an active r-mode pulsation balances the accretion torque
in this system.  This $L_q$ corresponds to an unabsorbed flux ($4\kpc$
distance) of $3\ee{-12}\erg\cm^{-2}\second^{-1}$, which is about ten times
the flux expected if the only heat source were crust nuclear reactions
\markcite{brown98:transients}({Brown} {et~al.} 1998).  Future \asca, \axaf, and \xmm\ observations
will assist in constraining the viscous damping present.  The luminosity
from the viscous damping is much larger than the expected magnetospheric
emission \markcite{becker97}({Becker} \& {Tr\"umper} 1997), and so interpretation of the spectrum should
be unambiguous in the absence of accretion onto the neutron star's
surface.

At $\timdot\lesssim 10^{-11}\Mspyr$, all of the viscous heating in the
core is radiated from the neutron star's surface during quiescence.  As
noted in section~\ref{sec:comp-observ-trans}, the relation between
$L_q(\timdot)$ then depends only on the accretion torque, and not on the
source distance and crust microphysics.  \axaf\ and \xmm\ are ideally
suited for a study of a population of low-luminosity neutron stars,
which offer excellent prospects for a clean determination of the amount
of viscous heating present.

At higher $\timdot$, for which neutrino cooling from the crust
contributes to balancing the viscous heating, the quiescent luminosity
depends on the crust microphysics.  In our calculations we assume that
the neutron crust is very impure and hence used thermal conductivity
dominated by electron-ion collisions.  If the conductivity of the crust
is higher than we have assumed (e.g., if the crust is a pure lattice),
then the predicted quiescent luminosity $L_q$ would be even higher than
that plotted in Figures~2 and~3.  In addition, we underestimated the
predicted $L_q$ by neglecting the effect of direct heating of the
neutron star crust by nuclear reactions occurring near neutron drip
\markcite{brown98:transients}({Brown} {et~al.} 1998).  Moreover, taking into account the
possibility that non-thermal emission contributes to the observed
quiescent luminosity further widens the gap between the observed $L_q$
and that inferred from the r-mode spin regulation hypothesis.  All of
these effects further strengthen our conclusions.

If the r-mode is not in steady state, then there remain several
possibilities: either the superfluid viscosity is so strong that it
suppresses the r-mode instability entirely, or the mode saturation
amplitude is so small that it is unimportant at all the spin frequencies
observed, or else the neutron star is in a limit cycle \markcite{levin99}({Levin} 1999)
of spin-up to some critical frequency, followed by rapid spin-down and
heating.  A detailed study of the spin evolution is necessary to
determine if the spin periods of the neutron stars are consistent with
such a scenario.  In particular, it remains an open question as to
whether one should expect to observe a population of slowly spinning
neutron stars with low-mass companions, such as Her~X-1, 4U~1626--67,
and GX~1+4.

The study of r-modes in neutron stars is rapidly evolving in response to
the interest aroused in the general relativity community.  While this
paper was being refereed, several theoretical developments occured that
are relevant for this study.  First, \markcite{lindblom99:superfluid}{Lindblom} \& {Mendell} (1999)
showed that unless the superfluid entrainment parameter assumes a very
special value, superfluid mutual friction is not competitive with
gravitational radiation for the r-mode amplitude evolution.  There is
therefore a conflict between theory and experiment: while theoretical
calculations show that r-modes in superfluid neutron stars should be
excited, the observations discussed in this paper are direct evidence
against the r-modes having a sufficient steady amplitude to limit the
spin of the neutron star, and the clustering of LMXB spin frequencies
argues against an recurrent instability.  This contradiction is likely
resolved by consideration of the presence of a solid crust
\markcite{bildsten99:_viscous_bound}(Bildsten \& Ushomirsky 1999), which dramatically enhances the
dissipation rate and damps the r-modes for typical core temperatures and
spin frequencies of LMXBs.  The findings presented in this paper lend
observational support to that conclusion.

\acknowledgements 

We thank Tom Prince for stimulating our interest in looking for
signatures of gravitational wave emission from LMXBs in ways that do not
require a gravitational wave detector.  We also thank Lars Bildsten,
Curt Cutler, Lee Lindblom, Ben Owen, and Yuri Levin for numerous
discussions.  This research was supported by NASA via grant NAGW-4517.
E.F.B is supported by a NASA GSRP Graduate Fellowship under grant
NGT5-50052.  G.U. acknowledges fellowship support from the Fannie and
John Hertz Foundation.


\end{document}